\documentclass[twocolumn,prd,amsmath,amssymb,floatfix]{revtex4}

\usepackage{bbold}
\usepackage{bm}
\usepackage{color}
\usepackage{dcolumn}
\usepackage{feynmf} 
\usepackage{graphics}
\usepackage{graphicx}
\usepackage{subfigure}
\usepackage{slashed}
\usepackage{placeins}

\usepackage{epsfig}

\def\al{\alpha}

\def\be{\begin{equation}}
\def\ee{\end{equation}}
\def\bea{\begin{eqnarray}}
\def\eea{\end{eqnarray}}
\def\bse{\begin{subequations}}
\def\ese{\end{subequations}}
\def\bc{\begin{center}}
\def\ec{\end{center}}

\def\ra{\rightarrow}

\def\nonum{\nonumber}

\def\D{{\rm d}}
\def\Ord{{\rm O}}

\def\Amu{{A_\mu}}

\def\Dmu{{\partial_\mu}}

\def\DDmu{{D_\mu}}

\begin{document}


\fmfcmd{%
vardef cross_bar (expr p, len, ang) =
((-len/2,0)--(len/2,0))
rotated (ang + angle direction length(p)/2 of p)
shifted point length(p)/2 of p
enddef;
style_def crossed expr p =
cdraw p;
ccutdraw cross_bar (p, 5mm,  45);
ccutdraw cross_bar (p, 5mm, -45)
enddef;}

\title{Critical behaviour of reduced QED$_{4,3}$ \\
and dynamical fermion gap generation in graphene 
}

\author{A.~V.~Kotikov$^1$ and S.~Teber$^{2,3}$}
\affiliation{
$^1$Bogoliubov Laboratory of Theoretical Physics, Joint Institute for Nuclear Research, 141980 Dubna, Russia.\\
$^2$Sorbonne Universit\'es, UPMC Universit\'e Paris 06, UMR 7589, LPTHE, F-75005, Paris, France.\\
$^3$CNRS, UMR 7589, LPTHE, F-75005, Paris, France.}

\date{\today}

\begin{abstract}
The dynamical generation of a fermion gap in graphene is studied at the 
infra-red Lorentz-invariant fixed point where the system is described
by an effective relativistic-like field theory: reduced QED$_{4,3}$ with $N$ four component fermions ($N=2$ for graphene), 
where photons are $(3+1)$-dimensional and mediate a fully retarded interaction among $(2+1)$-dimensional fermions.
A correspondence between reduced QED$_{4,3}$ and QED$_3$ allows us to derive an exact gap equation for QED$_{4,3}$ up to next-to-leading order.
Our results show that a dynamical gap is generated for $\alpha > \alpha_c$ where $1.03 < \alpha_c < 1.08$ in the case $N=2$
or for $N < N_c$ where $N_c$ is such that $\alpha_c \to \infty$ and takes the values $3.24 < N_c < 3.36$. 
The striking feature of these results is that they are in good agreement with values found in models with instantaneous Coulomb interaction (including lattice simulations).
At the fixed point: $\alpha = 1/137 \ll \alpha_c$, and the system is therefore in the semi-metallic regime in accordance with experiments.
\end{abstract}

\maketitle

\section{Introduction} 

Effects of electron-electron interactions in graphene, a one-atom thick layer of graphite, have been the subject of extensive experimental and theoretical works
for the last decade, see \cite{KotovUPGC12} for a review. In contrast to Galilean invariant band metals and semi-conductors, 
intrinsic (undoped) graphene is characterized by a massless Dirac-like quasiparticle spectrum at low-energies \cite{Semenoff84+Wallace47}. 
Because of the vanishing density of states at the two stable Fermi (or Dirac) points, the Coulomb interaction is poorly screened and remains long-ranged.
Moreover, the (bare) Fermi velocity is much smaller than the velocity of light: $v = c/300$, and, consequently, the (bare) fine structure
constant of graphene is of order $1$ in general: $\alpha= \al_{\text{QED}}\,c/(\kappa v) \approx 2.2/\kappa$
where $\al_{\text{QED}} = e^2/(4 \pi \hbar c)=1/137$. Suspended graphene, for which the dielectric constant $\kappa \approx 1$, 
is therefore often thought of as the condensed matter realization of strongly coupled quantum electrodynamics (QED) albeit non-relativistic.
Nevertheless, an early renormalization group approach revealed the existence of an
infrared Lorentz invariant fixed point for graphene \cite{GonzalezGV93}. Approaching this fixed point, the Fermi velocity flows to the velocity of light, $v \ra c$,
while the fine structure constant of graphene flows to the usual QED one: $\al \ra \al_{\text{QED}}$.
Experimentally, the flow of the velocity seems to be cut at values of the order of $3v$ in clean suspended graphene \cite{Eliasetal11} which might be interpreted as a slight effective
reduction of the coupling constant to $0.73$ which is still close to unity.

A commonly raised theoretical issue is then to know whether or not the strong and long-ranged Coulomb interaction may drive the semi-metallic 
system into an insulating phase thereby dynamically generating a gap in the single particle fermionic spectrum, see, {\it e.g.},
 \cite{Khveshchenko01,GorbarGMS02,LealK03,Son07,Khveshchenko08,LiuLC09,GamayunGG09,Drut08,Gonzalez11,WangL12,BuividovichP13,Popovici13,Gonzalez15,Katanin15}.
This excitonic instability should take place for values of the coupling constant larger than a critical one, $\al_c$, {\it i.e.}, a gap is dynamically generated for
$\al > \al_c$ for the physical number of fermion flavours (spin): $N=2$. Alternatively, the instability would also manifest at $N <N_c$ where 
$N_c$ is a critical fermion flavour number for which $\al_c \to \infty$. A quantitative analysis of such an instability is important given the fact that
the possibility to generate in a controlled way a fermion gap in graphene and graphene-like materials is crucial for, {\it e.g.}, the development of graphene-based transistors \cite{Neviusetal15}.
Moreover, such an instability is the condensed matter physics analogue of the much studied dynamical mass generation and associated dynamical
chiral symmetry breaking (D$\chi$SB) in usual relativistic QEDs and in particular in $(2+1)$-dimensional QED or QED$_3$ see, {\it e.g.}, 
\cite{Pisarski84,AppelquistNW88,Nash89,Kotikov93+12,KotikovST16,Giombi:2016fct,Herbut16,Gusynin:2016som,Kotikov:2016prf}. Of theoretical importance, is to
clarify the precise relation between QED$_3$ and effective field theories describing graphene.

A puzzling fact is that, despite the strength of the interaction, there is no experimental evidence for the existence of a gap of more than $0.1$meV
in clean suspended graphene \cite{Eliasetal11}. This may be the indication that $\al$ is not large enough in actual samples and/or that the latter may be subject to additional sources of screening,
{\it e.g.}, from electrons in $\sigma$ bands \cite{UlybyshevBKP13}. Neglecting the latter and focusing on clean graphene at zero temperature, there have been many attempts 
to compute $\al_c$ on the basis of an elementary model 
of massless Dirac fermions interacting via the instantaneous Coulomb interaction, {\it i.e.}, the limit $v/c \ra 0$ which is indeed quite realistic. 
Various methods were used such as, {\it e.g.}, analytical or numerical solutions of leading order (LO) Schwinger-Dyson (SD) equations 
\cite{Khveshchenko01,GorbarGMS02,LealK03,Khveshchenko08,LiuLC09,GamayunGG09,WangL12,Popovici13,Gonzalez15}, renormalization group studies \cite{Son07,Gonzalez11}, 
lattice simulations \cite{Drut08,BuividovichP13} and a combination of Bethe-Salpeter and functional renormalization group approaches \cite{Katanin15}.
All computations agree on the fact that $\al_c \sim \Ord(1)$ but there is still
no general agreement on the precise value of $\al_c$ 
even though recent results \cite{WangL12,Popovici13,Gonzalez15,Katanin15} seem to indicate that it is indeed larger than the bare value of $\al$ in agreement with the experimentally observed semi-metallic behaviour.

In the present paper, we revisit the problem of dynamical gap generation in graphene from a completely different angle.
We shall focus on the deep infra-red Lorentz invariant fixed point of \cite{GonzalezGV93} where graphene 
may be effectively described by a massless relativistic quantum field theory 
model whereby $d_e=2+1$-dimensional electrons interact via a $d_\gamma=3+1$-dimensional long-range fully retarded potential, {\it i.e.}, the limit $v \ra c$.
As described in Sec.~\ref{sec:model}, such a model belongs to the class of reduced QED \cite{GorbarGM01}, see also \cite{Marino93+DoreyM92+KovnerR90}, 
or RQED$_{d_\gamma,d_e}$, and corresponds to RQED$_{4,3}$ in the case of graphene \cite{Teber12+KotikovT13,KotikovT14}.
Our results will provide an exact analytical solution for the SD equations of RQED$_{4,3}$ up to next-to-leading order (NLO) including a full resummation of the wave function renormalization constant and a proof of the
weak gauge-variance of $\al_c$ and $N_c$.
They are based on an important correspondence between RQED$_{4,3}$ and large-$N$ QED$_3$ which is described in Sec.~\ref{sec:cor}. The latter will allow us, in Secs.~\ref{sec:LO} and \ref{sec:NLO},
to transcribe the recent exact NLO gap equation found for QED$_3$ \cite{Kotikov:2016prf} to the case of RQED$_{4,3}$. 
Solving the gap equation will yield high precision estimates of $\al_c$ and $N_c$. Comparison with other results that can be found in the literature 
will be given in Sec.~\ref{sec:compare} and the conclusion in Sec.~\ref{sec:conclude}.
 In the following, we work in units in which $\hbar = c = 1$.

\section{The model}
\label{sec:model}

 The effective low-energy action describing graphene at the Lorentz invariant fixed point is that 
of massless RQED$_{4,3}$ which is given by:
\bea
S = \sum_{\sigma=1}^{N} \int \D^3 x \, \overline{\Psi}_\sigma i \gamma^{\mu}D_{\mu} \Psi_\sigma - \int \D^4 x \bigg(\frac{F_{\mu \nu}^2}{4} + \frac{\left(\partial_\mu A^{\mu}\right)^2}{2\overline{\xi}} \bigg)\, ,
\nonum
\eea
where $\Psi$ is a four-component spinor with $N$ flavours ($N=2$ for graphene), $\gamma^\mu$ is the  Dirac matrix satisfying the
usual algebra: $\{ \gamma^\mu,\gamma^\nu \} = 2 g^{\mu \nu}$ where $g^{\mu \nu} = {\rm diag}(1,-1,-1)$ is the metric tensor, 
$\DDmu = \Dmu + ie \Amu$ is the covariant derivative and $\bar{\xi} = 1 - \bar{\eta}$ is the gauge fixing parameter associated with
the $(3+1)$-dimensional gauge field. Similarly to QED$_3$, this actions has a $U(2N)$ flavour symmetry. Dynamical mass generation
leads to a spontaneous breakdown of this symmetry to $U(N) \times U(N)$. Notice, however, that while QED$_3$ is super-renormalizable
and has an intrinsic mass scale fixed by the dimensionful coupling constant $a =N e^2/8$ \cite{Pisarski84}, RQED$_{4,3}$ is renormalizable
and has a dimensionless coupling $\al = e^2/(4\pi)$ \cite{GorbarGM01,Teber12+KotikovT13,KotikovT14}. As noticed in \cite{GorbarGM01}, D$\chi$SB
in RQED$_{4,3}$ corresponds to the so-called conformal phase transition (CPT) whereas it is only a pseudo-CPT in QED$_3$ \cite{MiranskyY97}.
Nevertheless, in both cases, the dynamical mass satisfies the Miransky scaling albeit with different coefficients in front of the exponential, see \cite{GorbarGM01}.
In the following, we shall not be interested by the full scaling but rather focus on the critical behaviour of RQED$_{4,3}$, {\it i.e.}, 
consider the transition point $\al=\al_c$ or $N = N_c$ where the dynamical mass first appears.

In order to study the dynamical generation of such a mass we need to solve the SD equations for the fermion propagator. 
As we have mentioned above, the NLO solution of such equations has recently been done in the case of QED$_3$
in the $1/N$-expansion \cite{Kotikov:2016prf}. Technically, the calculations were performed with the help of the standard rules of perturbation theory for massless Feynman diagrams
as in [\onlinecite{Kazakov83}] and with the help of the Gegenbauer polynomial technique as used in \cite{Kotikov95}, see also the recent short review [\onlinecite{TeberK16}].
Similar calculations can be carried out for RQED$_{4,3}$ on the basis of the multi-loop results obtained in \cite{Teber12+KotikovT13,KotikovT14}.
An alternative way to derive the same results is based on a 
simple correspondence between RQED$_{4,3}$ and large-$N$ QED$_3$. As we shall demonstrate in the following, the latter will allow us to straightforwardly study the critical properties of RQED$_{4,3}$ without any 
further complicated calculation.

\section{Correspondence between reduced QED$_{4,3}$ and QED$_3$} 
\label{sec:cor}

We first consider the photon propagator 
of RQED$_{4,3}$ \cite{Teber12+KotikovT13,KotikovT14} (in Euclidean space):
\begin{flalign}
D_{\text{RQED}}^{\mu \nu}(p)
= \frac{1}{2|p|} d^{\mu \nu}\left(\frac{\overline{\eta}}{2}\right),~~~
d^{\mu \nu}(\eta) = g^{\mu \nu} - \eta \frac{p^{\mu}p^{\nu}}{p^2}\, . 
\label{D-RQED}
\end{flalign}
From Eq.~(\ref{D-RQED}) we see that we may define an effective gauge fixing parameter, $\eta = 1 - \xi$, 
for the reduced gauge field which is related to the gauge fixing parameter, $\overline{\eta} = 1-\overline{\xi}$, 
of the corresponding $4$-dimensional gauge field as follows:
\be
\eta= \frac{\overline{\eta}}{2}, \qquad \xi = \frac{1+\overline{\xi}}{2}\, .
\label{defxi}
\ee

Next, we consider the photon propagator 
of QED$_{3}$ 
in a non-local $\xi$-gauge \cite{Nash89}:
\be
D^{\mu \nu}_{\text{QED}3}(p) = 
\frac{d^{\mu \nu}(\eta)}{p^2 \left[1 + \Pi (p) \right]} \, ,
\label{D-QED3}
\ee
where $\Pi(p)$ is the polarization operator.
At the LO of the $1/N$-expansion, the polarization operator reads:
\be
\Pi_1(p) = \frac{a}{|p|}\, .
\label{Pi1}
\ee
In the large-$N$ limit, the infrared behaviour of $D^{\mu \nu}_{\text{QED}3}(p)$ changes \cite{Kotikov93+12,KotikovST16}, see also \cite{AppelquistP81,JackiwT81+AppelquistH81}, 
because: 
\be
D^{\mu \nu}_{\text{QED}3}(p) = \frac{8}{e^2 N |p|} d^{\mu \nu}(\eta) \, .
\label{D-QED3-N}
\ee
As already noticed in \cite{Teber12+KotikovT13,KotikovT14} the photon propagators of QED$_3$ in the large-$N$ limit, Eq.~(\ref{D-QED3-N}), and the
one of RQED$_{4,3}$, Eq.~(\ref{D-RQED}) have the same form. One may easily pass from one form to the other with the help of the following transformation:
%
%
\be
\frac{1}{L} \to \frac{e^2}{16\pi^2} = \frac{\al}{4\pi} \equiv g, \quad \eta \to \frac{\overline{\eta}}{2} \quad \left( \xi \to \frac{1+\overline{\xi}}{2} \right) \, ,
\label{transform.1}
\ee
where $L=\pi^2 N$ \cite{KotikovST16,Kotikov:2016prf}.
The transformation (\ref{transform.1}) will allow us to transcribe the recent solution
of dynamic mass generation in QED$_3$ using the $1/N$-expansion \cite{KotikovST16,Gusynin:2016som,Kotikov:2016prf}
to the case of RQED$_{4,3}$ using the loop expansion. In the following, LO will either refer to LO 
in the $1/N$-expansion for QED$_3$ or to the one-loop order for RQED$_{4,3}$. Similarly, NLO will either refer
to NLO in the $1/N$-expansion for QED$_3$ or to the two-loop order for RQED$_{4,3}$. Of course, the solution of SD
equations, combined with the various resummations we shall perform in the following, is non-perturbative in nature
and beyond the reach of a simple $1/N$ or loop expansion. 

%
%

\section{Leading order} 
\label{sec:LO}

In order to illustrate how the correspondence works, we first compute the critical coupling at LO.
Combining the LO QED$_3$ result \cite{Kotikov:2016prf} with the transformation (\ref{transform.1}), the LO gap equation for the critical coupling of RQED$_{4,3}$ reads:
\be
1 = \frac{16(2+\xi)}{L_c} \quad \to \quad 1 = 16(2+\xi)g_c = 8(5+ \overline{\xi}) g_c \, ,
\label{GE-LO}
\ee
where $g_c = \al_c/(4\pi)$. This yields:
\be
\al_c(\overline{\xi}) = \frac{\pi}{2(5 + \overline{\xi})}\, .
\label{alc-LO}
\ee
The LO critical coupling is seen to be strongly gauge dependent but does not depend on the fermion flavour number, $N$. 
Its value in various gauges, including Landau ($\overline{\xi}=0,\xi=-1$) and Feynman ($\overline{\xi}=\xi=1$) gauges, 
reads:
\begin{subequations}
\label{alc-LO-values}
\begin{flalign}
&\alpha_c (\overline{\xi}=0)=0.3142, \quad ~~ \alpha_c(\overline{\xi}=1) =0.2618\, ,
\label{alc-LO-values1}\\
&\alpha_c(\overline{\xi}=-1) =0.3927, \quad \alpha_c(\overline{\xi}=1/3) = 0.2945\, .
\label{alc-LO-values2}
\end{flalign}
\end{subequations}
The gauge ($\overline{\xi}=-1,\xi=0$) corresponds to the Landau gauge for the reduced gauge field while 
the gauge ($\overline{\xi}=1/3,\xi=2/3$) will be discussed later.
 
Following, {\it e.g.}, \cite{GorbarGMS02,Khveshchenko08,LiuLC09} (see also the review \cite{KotovUPGC12})
the dynamical screening of the interaction may be included, in the so-called random-phase approximation (RPA), 
by resumming the one-loop polarization operator, Eq.~(\ref{Pi1}); notice that the LO polarization operator in QED$_3$ and the one-loop
polarization operator in RQED$_{4,3}$ are equal.
Contrarily to the case of QED$_{3}$, however, in RQED$_{4,3}$ such a resummation does not change the infrared property of the corresponding photon propagator
\cite{Teber12+KotikovT13} which then reads:
\be
D_{RQED}^{\mu \nu}(p) =
\frac{1}{2|p|(1+N e^2/16)} d^{\mu \nu}\left(\frac{\tilde{\eta}}{2}\right) \, .
\label{D-RQED-RPA}
\ee
This RPA resummation may be taken into account with the help of a simple redefinition of the coupling constant which becomes:
\be
\tilde{\al} = \frac{\al}{1+Ne^2/16}\, .
\label{al-RPA}
\ee
Accordingly, the transformation (\ref{transform.1}) has to be replaced by:
\be
\frac{1}{L} \to \frac{\tilde{e}^2}{16\pi^2} = \frac{\tilde{\al}}{4\pi} \equiv \tilde{g}, \quad \eta \to \frac{\overline{\eta}}{2} \quad \left( \xi \to \frac{1+\overline{\xi}}{2} \right) \, .
\label{transform.2}
\ee
The gap equation (\ref{GE-LO}) then immediately yields the LO RPA critical coupling constant:
\be
\al_c(\overline{\xi}) = \frac{\pi}{2(5+\overline{\xi}) - N \pi^2/4}\, ,
\label{alc-LO+RPA}
\ee
which now is not only gauge dependent but also depends on the number of fermion flavours, $N$.
For $N=2$, Eq.~(\ref{alc-LO+RPA}) yields the following values in various gauges:
\begin{subequations}
\label{alc-LO+RPA-values}
\begin{flalign}
&\alpha_c (\overline{\xi}=0) = 0.6202, \quad ~~ \alpha_c(\overline{\xi}=1) = 0.4447\, ,
\label{alc-LO+RPA-values1}\\
&\alpha_c(\overline{\xi}=-1) = 1.0249, \quad \alpha_c(\overline{\xi}=1/3) = 0.5481\, .
\label{alc-LO+RPA-values2}
\end{flalign}
\end{subequations}
Dynamical screening therefore increases the value of the critical coupling, compare (\ref{alc-LO+RPA-values}) with (\ref{alc-LO-values}).
It is also convenient to find the critical fermion flavour number $N_c$ for which $\al_c \to \infty$.
Eq.~(\ref{alc-LO+RPA}) yields:
\be
N_c(\overline{\xi}) = \frac{8(5 + \overline{\xi})}{\pi^2}\, .
\label{Nc-LO}
\ee
This number coincides 
 with the critical flavour number $N_c$ for D$\chi$SB in QED$_3$ which is defined as $N_c=L_c/\pi^2$.
In various gauges, its value reads:
\begin{subequations}
\label{Nc-LO+RPA-values}
\begin{flalign}
&N_c(\overline{\xi}=0) = 4.0529, \quad ~~ N_c(\overline{\xi}=1) = 4.8634\, ,
\label{Nc-LO+RPA-values1}\\
&N_c(\overline{\xi}=-1) = 3.2423, \quad N_c(\overline{\xi}=1/3) = 4.3230\, .
\label{Nc-LO+RPA-values2}
\end{flalign}
\end{subequations}
Notice that the value $N_c(1/3) = 128/(3\pi^2)$ has already been obtained in \cite{GorbarGM01} where the importance of the 
$\overline{\xi}=1/3$ gauge has been emphasized following the seminal work of Nash \cite{Nash89}, see also discussions in \cite{Kotikov:2016prf}. 
As shown by Nash \cite{Nash89}, the wave function renormalization constant may be resummed at the level of the gap equation.
The peculiar $\overline{\xi}=1/3$ gauge is the one where the wave function renormalization constant vanishes at LO and for which Nash's 
resummation does not affect much the results. We shall confirm this below.

\section{Next-to-Leading order} 
\label{sec:NLO}

From our NLO results for QED$_3$ \cite{Kotikov:2016prf}, combined with the transformations (\ref{transform.1}) 
or (\ref{transform.2}), we may compute the NLO coupling constant of RQED$_{4,3}$. However, in order to properly do so, some additional
replacements are necessary.

The first additional replacement is related to the fact that, in QED$_3$, the NLO polarization operator $\hat{\Pi}_2$ \cite{Kotikov:2016prf} 
contributed to the NLO result within the framework of the $1/N$ expansion:
\be
\Pi_2^{(\text{QED$_3$})}(p) = \frac{2a}{N\pi^2}\,\frac{\hat{\Pi}_2}{|p|}, \quad \hat{\Pi}_2 = \frac{92}{9} - \pi^2 \, .
\label{Pi2-QED3}
\ee
On the other hand, within the framework of standard perturbation theory which applies to RQED$_{4,3}$, it is only the LO (one-loop) polarization operator, Eq.~(\ref{Pi1}), which contributes at NLO (two-loop).
From the ratio of Eqs.~(\ref{Pi2-QED3}) and (\ref{Pi1}) we define $\hat{\Pi}_1$ which is such that:
\be
\frac{\Pi_2^{(\text{QED$_3$})}(p)}{\Pi_1(p)} = \frac{\hat{\Pi}_2}{\hat{\Pi}_1}, \quad \hat{\Pi}_1 = \frac{N \pi^2}{2} \, .
\label{hatPi1}
\ee
So, in going from large-$N$ QED$_3$ to RQED$_{4,3}$, the first additional replacement reads:
\be
\hat{\Pi}_2  \quad  \to \quad \hat{\Pi}_1 \, .
\label{transform.3}
\ee
As we shall see below, of importance will be the fact that the values of $\hat{\Pi}_1$ are large in comparison with those of $\hat{\Pi}_2$. 

The second additional replacement is related to the use of the non-local $\xi$-gauge in the case of QED$_{3}$. Transforming back to the usual
$\xi$-gauge in RQED$_{4,3}$ amounts to apply the following simple condition: 
\be
\xi \hat{\Pi}_1 =  0\, .
\label{transform.4}
\ee

With these additional transformations we are now in a position to transcribe the NLO results of large-$N$ QED$_3$ to the case of RQED$_{4,3}$.
In the following, we shall first solve the NLO gap equation without Nash's resummation and then with Nash's resummation.

\subsection{NLO without Nash's resummation} 

Combining the QED$_3$ NLO results of  \cite{Kotikov:2016prf} with the transformations (\ref{transform.1}), (\ref{transform.3}) and (\ref{transform.4}), the NLO gap equation 
for the critical coupling of RQED$_{4,3}$ without RPA resummation reads: 
\begin{flalign}
&1 = 16(2+\xi) g_c
\nonum \\
&- 8\left(S(\xi)- 16 \left(4-\frac{50}{3}\xi + 5\xi^2\right) - 8 \hat{\Pi}_1 \right)\,g_c^2 \, ,
\label{GE-NLO1}
\end{flalign}
where we have kept $\xi = (1+\overline{\xi})/2$ to facilitate comparison with the results of \cite{Kotikov:2016prf}.
In Eq.~(\ref{GE-NLO1}), $S(\xi)$ contains the contribution of the complicated diagrams $I_1$, $I_2$ and $I_3$
having representations \cite{KotikovST16} in the form of two-fold series (for $I_1$) and three-fold ones (for $I_2$ and $I_3$),
respectively. Solving Eq.~(\ref{GE-NLO1}), we have two standard solutions:
%
\begin{flalign}
\al_{c,\pm}(\xi) = \frac{4\pi}{8(2+\xi) \pm \sqrt{d_1(\xi)}}\, ,
\label{alc-NLO1}
%
\end{flalign}
%
where 
\begin{flalign}
&d_1(\xi)= 8\left(S(\xi)-8\left(4-\frac{112}{3}\xi+9\xi^2\right)- 8
\hat{\Pi}_1 \right) \, .
\label{d1}
\end{flalign}
It turns out that the ``$-$'' solution is unphysical and has to be rejected because $\al_{c,-}<0$. So, the physical solution is unique and corresponds to
$\al_{c} = \al_{c,+}$. For numerical applications, we use the numerical estimates of \cite{KotikovST16,Kotikov:2016prf}:
%
\begin{flalign}
R_1=163.7428, ~~ R_2=209.175, ~~ P_2=1260.720 \, ,
\label{R1.R2.P2-values} 
\end{flalign}
%
which enter the expression of $S(\xi)$:
\begin{flalign}
S(\xi) =
(1-\xi)R_1
- (1-\xi^2) \frac{R_2}{8} - (7+16 \xi -3 \xi^2) \frac{P_2}{128}\, .
\label{S-expr}
\end{flalign}
%
In various $\overline{\xi}$-gauges: $\overline{\xi}=0,1,-1,1/3$ that respectively correspond to $\xi=1/2,1,0,2/3$, 
Eq.~(\ref{R1.R2.P2-values}) allows us to obtain the numerical value of: 
%
\begin{subequations}
\label{S-xi-values}
\begin{flalign}
&S(\overline{\xi}=0)=\frac{R_1}{2}-\frac{3 R_2}{32} - \frac{57 P_2}{512}\, ,
\label{S-oxi=0} \\
&S(\overline{\xi}=1)= -\frac{5P_2}{32}\, ,
\label{S-oxi=1} \\
&S(\overline{\xi}=-1)=R_1-\frac{R_2}{8}-\frac{7P_2}{128}\, ,
\label{S-oxi=-1} \\ 
&S(\overline{\xi}=1/3)=\frac{R_1}{3}-\frac{5 R_2}{72} - \frac{49 P_2}{384}\, .
\label{S-oxi=1/3}
\end{flalign}
\end{subequations}
Notice that the solutions of Eq.~(\ref{alc-NLO1}) are physical provided that the following inequality is satisfied: 
\be
d_1(\xi) \geq 0 \, .
\label{inequality.1}
\ee
In the absence of RPA resummation, Eq.~(\ref{inequality.1}) 
is satisfied only in the unphysical case $N=0$ for which $\hat{\Pi}_1=0$.
For $N>0$, the large value of $\hat{\Pi}_1$ makes $d_1(\xi)$ negative which in turn implies that Eq.~(\ref{GE-NLO1}) has no physical solutions.

Fortunately, the situation strongly improves upon performing the RPA resummation.
%
Indeed, in this case, the inequality Eq.~(\ref{inequality.1}) 
is satisfied for all the values of $\xi$ we consider, excepting the case $\overline{\xi}=\xi=1$ which corresponds to the Feynman gauge.
To see this, consider Eq.~(\ref{al-RPA}) which can be re-written as:
\be
\al = \frac{\tilde{\al}}{1 - \hat{\Pi}_1 \tilde{\al}/(2\pi)}, \quad g = \frac{\tilde{g}}{1 - 2\hat{\Pi}_1 \tilde{g}}\, ,
\label{al.vs.al-RPA}
\ee
where $\hat{\Pi}_1$ defined in Eq.~(\ref{hatPi1}) was made explicit. Substituting Eq.~(\ref{al.vs.al-RPA}) in (\ref{GE-NLO1}), 
we see that the contribution of $\hat{\Pi}_1$ cancels out from the new gap equation which reads:
\begin{flalign}
1 = 16(2+\xi) \tilde{g}_c
- 8\left(S(\xi)- 16 \left(4-\frac{50}{3}\xi + 5\xi^2\right) \right)\,\tilde{g}_c^2 \, .
\label{GE-NLO2}
\end{flalign}
Hence, Eq.~(\ref{GE-NLO2}) has a broader range of solutions than Eq.~(\ref{GE-NLO1}) including
all $\overline{\xi}$-values such as: $-5.695 < \overline{\xi} < 0.915$ or, equivalently: 
$-2.348 < \xi < 0.957$; the Feynman gauge is still excluded because $S(\overline{\xi}=1)$ is large and negative. 
Numerical applications for $\tilde{\al}_c = \tilde{\al}_{c,+}$ then yield:
%
\be
\tilde{\al}_c(0)= 0.3804,~\tilde{\al}_c(-1)= 0.3794,~ \tilde{\al}_c(1/3)= 0.3924\, .
\label{talc-NLO2}
\ee
%
%
Substituting the values (\ref{talc-NLO2}) in Eq.~(\ref{al.vs.al-RPA}) yields, for $N=2$:
%
\be
\al_c(0)= 0.9451, ~\al_c(-1)= 0.9389, ~\al_c(1/3)= 1.0227\, .
\ee

The critical number, $N_c$, for which $\al_c \to \infty$ and is such that a finite critical
coupling exists for $N<N_c$ takes the following values:
%
\be
N_c(0)= 3.3472, ~N_c(-1)= 3.3561, ~ N_c(1/3)= 3.2450\, .
\label{Nc-NLO2}
\ee
%
%
%
%
%
%
%
%
The numerical values in (\ref{Nc-NLO2}) are a little larger than in QED$_3$ where $N_c$ is defined as $N_c=L_c/\pi^2$ because of the additional
(small) factor $\hat{\Pi}_2$ coming with a negative sign, see \cite{Kotikov:2016prf}.

%
%
%

\subsection{NLO with Nash's resummation}  

Following \cite{Nash89} we now resum the ``basic'' part of the NLO corrections corresponding to the fermion wave function renormalization.
Combining the QED$_3$ resummed gap equation of \cite{Kotikov:2016prf} with the transformations (\ref{transform.1}), (\ref{transform.3}) and (\ref{transform.4}), the NLO resummed gap equation
for the critical coupling of RQED$_{4,3}$ without RPA resummation reads:
\be
1 = \frac{128}{3} g_c + 8 \left( \tilde{S}(\xi) - \frac{1280}{27} - \frac{32}{3} \hat{\Pi}_1 \right)g_c^2 \, ,
%
\label{GE-NLO3} 
\ee
where $\tilde{S}(\xi)$ contains the rest of $S(\xi)$, see \cite{Kotikov:2016prf}, after the extraction of the ``most important''
contributions.
%
Similarly to the case of QED$_3$ \cite{Kotikov:2016prf}, the striking feature of Eq.~(\ref{GE-NLO3}) is the absence of $\xi$-dependence at LO
and it's strong suppression at NLO: the $\xi$-dependence does exist at NLO but only via $\tilde{S}(\xi)$ which, as we shall see shortly, is small numerically.
Solving Eq.~(\ref{GE-NLO3}), we have the two standard solutions:
\be
\al_{c,\pm} = \frac{4\pi}{64/3 \pm \sqrt{d_2(\xi)}}\, ,
\label{alc-NLO3}
\ee
where 
\be
d_2(\xi)= 8 \left( \tilde{S}(\xi) + \frac{256}{27} - \frac{32}{3} \hat{\Pi}_1 \right)\, .
\label{d2}
\ee
%
%
%
As before, the ``$-$'' solution is unphysical and has to be rejected because $\al_{c,-}<0$. So, the physical solution is unique and corresponds to
$\al_{c} = \al_{c,+}$. In order to provide  numerical estimates,
we use the values $\tilde{R}_1$, $\tilde{R}_2$ and $\tilde{P}_2$, see \cite{Kotikov:2016prf}:
\be
\tilde{R}_1=3.7428, \quad \tilde{R}_2=1.175, \quad \tilde{P}_2=-19.280 \, ,
\label{tIs-numerics}
\ee
which enter the expression of $\tilde{S}(\xi)$:
\be
\tilde{S}(\xi) = 
(1-\xi)\tilde{R}_1
- (1-\xi^2) \frac{\tilde{R}_2}{8} - (7+16 \xi -3 \xi^2) \frac{\tilde{P}_1}{128}\, .
\label{sigma2.3} 
\ee
For the gauge choices: $\overline{\xi}=0,1,-1,1/3$ that respectively correspond to $\xi=1/2,1,0,2/3$,
we may then obtain numerical values of: 
%
\begin{subequations}
\label{tS-xi-values}
\begin{flalign}
&S(\overline{\xi}=0)= \frac{\tilde{R}_1}{2}-\frac{3 \tilde{R}_2}{32} - \frac{57 \tilde{P}_2}{512}\, ,
\label{tS-oxi=0} \\
&S(\overline{\xi}=1)= -\frac{5\tilde{P}_2}{32}\, ,
\label{tS-oxi=1} \\
&S(\overline{\xi}=-1)= \tilde{R}_1-\frac{\tilde{R}_2}{8}-\frac{7\tilde{P}_2}{128}\, ,
\label{tS-oxi=-1} \\
&S(\overline{\xi}=1/3)= \frac{\tilde{R}_1}{3}-\frac{5 \tilde{R}_2}{72} - \frac{49 \tilde{P}_2}{384}\, .
\label{tS-oxi=1/3}
\end{flalign}
\end{subequations}
%
%
%
Notice that the solutions of Eq.~(\ref{alc-NLO3}) are physical provided that: 
\be
d_2(\xi) \geq 0
\label{inequality.2}
\ee
As in the previous case, this inequality is satisfied only for the nonphysical case $N=0$.

Fortunately, the situation once again strongly improves upon the additional implementation
of the RPA resummation. Substituting Eq.~(\ref{al.vs.al-RPA}) in (\ref{GE-NLO3}),
we see that the contribution of $\hat{\Pi}_1$ cancels out from the new gap equation which reads:
\be
1 = \frac{128}{3} \tilde{g}_c + 8 \left( \tilde{S}(\xi) - \frac{1280}{27} \right) \tilde{g}_c^2 \, .
\label{GE-NLO4}
\ee
Hence, Eq.~(\ref{GE-NLO4}) has a broader range of solutions than Eq.~(\ref{GE-NLO3}) including
all $\overline{\xi}$-values such as: $-19.668 < \overline{\xi} < 8.928$ or, equivalently:
$-9.334 < \xi < 4.964$. This shows that the improvement is even better that in the absence of Nash's resummation because,
in the present case, a physical solution also exists in the Feynman gauge $\overline{\xi}=\xi=1$. 
Numerical applications for $\tilde{\al}_c = \tilde{\al}_{c,+}$ yield:
\begin{subequations}
\label{talc-NLO4}
\begin{flalign}
&\tilde{\al}_c(\overline{\xi}=0)= 0.3966, \quad ~~ \tilde{\al}_c(\overline{\xi}=1)= 0.4011\, ,
\\
&\tilde{\al}_c(\overline{\xi}=-1)= 0.3931, \quad \tilde{\al}_c(\overline{\xi}=1/3)= 0.3980\, .
\end{flalign}
\end{subequations}
Substituting the values (\ref{talc-NLO4}) in Eq.~(\ref{al.vs.al-RPA}) yields, for $N=2$:
\begin{subequations}
\label{alc-NLO4-N=2}
\begin{flalign}
&\al_c(\overline{\xi}=0)= 1.0521, \quad ~~ \tilde{\al}_c(\overline{\xi}=1)= 1.0841\, ,
\\
&\al_c(\overline{\xi}=-1)= 1.0278, \quad \al_c(\overline{\xi}=1/3)= 1.0619\, .
\end{flalign}
\end{subequations}
These values are very close to each other proving the very weak gauge variance of our results.
For the sake of completeness, we give the value of the critical coupling constant in the case $N=1$:
%
\begin{subequations}
\label{alc-NLO4-N=1}
\begin{flalign}
&\al_c(\overline{\xi}=0)= 0.5761, \quad ~~ \tilde{\al}_c(\overline{\xi}=1)= 0.5855\, ,
\\
&\al_c(\overline{\xi}=-1)= 0.5687, \quad \al_c(\overline{\xi}=1/3)= 0.5790\, ,
\end{flalign}
\end{subequations}
and in the case $N=3$:
%
\begin{subequations}
\label{alc-NLO4-N=3}
\begin{flalign}
&\al_c(\overline{\xi}=0)= 6.0588, \quad ~~ \tilde{\al}_c(\overline{\xi}=1)= 7.2971\, ,
\\
&\al_c(\overline{\xi}=-1)= 5.3310, \quad \al_c(\overline{\xi}=1/3)= 6.3960\, .
\end{flalign}
\end{subequations}
For higher (integer) values of $N$, there is no instability. This can be seen by computing the value
of $N$, $N_c$, for which $\al_c \to \infty$. As in the previous case, see Eq.~(\ref{Nc-NLO2}),
this value 
coincides with the critical value $N_c$ for D$\chi$SB in QED$_3$:
\begin{subequations}
\label{Nc-NLO4}
\begin{flalign}
&N_c(\overline{\xi}=0)= 3.2102, \quad ~~ N_c(\overline{\xi}=1)= 3.1745\, ,
\\
&N_c(\overline{\xi}=-1)= 3.2388, \quad N_c(\overline{\xi}=1/3)= 3.1991\, .
\end{flalign}
\end{subequations}
As anticipated above, we see that the ``right(est)'' gauge choice \cite{GorbarGM01} is the one close to $(\overline{\xi}=1/3,\xi=2/3)$ where
the results are more or less the same before and after Nash's resummation. 

At this point, we would like to remark that the weakness of the gauge dependence 
of our results makes it unimportant from the point of view of physical applications. Moreover, such gauge dependence is not specific to reduced QED; actually, as the
mapping we have used suggests, it originates from a similar feature taking place in QED$_3$, see \cite{Kotikov:2016prf}. 
Nevertheless, the existence of such a gauge dependence, even though very weak, may call into question the applicability of our approach.
It is indeed well known that gauge dependence does not affect the critical value of $N_c$ in the case of QED$_3$, see \cite{BashirRSR09,Bashir:2008fk},
a statement which is based on an application of the Landau-Khalatnikov-Fradkin (LKF) transformation \cite{LandauK55+Fradkin56} to QED$_3$. 
Recently, an application of the LKF transformation has been carried out in the case of reduced QED
\cite{AhmadCCR16}. Note, however, that in our study of QED$_3$, \cite{Kotikov:2016prf}, we have worked in a non-local gauge
which is quite popular now in the $3$-dimensional case. In this case, a direct application of the LKF
transformation is quite problematic.  So, the study of a (non-local) gauge dependence needs additional
investigations and we hope to return to this problem in our future studies.

We finally consider the case where $\tilde{S}(\xi)=0$ in (\ref{GE-NLO4}).
In this case, there no
gauge dependence at all and we have: $\tilde{\al}_c = 0.41828$.
%
%
Hence: 
\begin{subequations}
\begin{flalign}
&\alpha_c(N=2)=1.2196 \, ,
\\
&\alpha_c(N=1)=0.6229, \quad \alpha_c(N=3)=28.9670\, .
\end{flalign}
\end{subequations}
The number $N_c$ coincides with $N_c=L_c/\pi^2$ in QED$_3$ and has the following value:
%
\be
\overline{N}_c=3.0440 \, ,
\ee
which is a little less than the ones in (\ref{Nc-NLO4}).



\section{Comparison with other results}
\label{sec:compare}

Our results for $\alpha_c$ ($0.94 < \alpha_c < 1.02$ without Nash's resummation and $1.03 < \alpha_c < 1.08$
with Nash's resummation) are in good agreement with $\alpha_c=0.92$ \cite{GamayunGG09} and $\alpha_c=1.13$ 
\cite{Khveshchenko08}. These last results were obtained as improvements of previous studies: \cite{GamayunGG09} took into account of the dynamical screening of
the interactions with respect to \cite{GorbarGMS02} where the value $\alpha_c=1.62$ was found in the static
approximation, {\it i.e.}, RPA with polarization operator at zero frequency; \cite{Khveshchenko08} took into account Fermi velocity renormalization
with respect to the earlier work in \cite{Khveshchenko01}; see also discussions in Refs.~\cite{GamayunGG09}
and \cite{Katanin15} as well as a detailed summary of these results in the review \cite{KotovUPGC12} and in \cite{WangL12}. 
Our results are also in good agreement with lattice Monte-Carlo simulations where  $\al_c = 1.11 \pm 0.06$ was obtained in \cite{Drut08} 
and $\al_c = 0.9 \pm 0.2$ in \cite{BuividovichP13}.
%
%
Moreover, in the strong coupling regime, $\alpha_c \to \infty$, our critical values for $N_c$ ($3.24 < N_c <  3.36$ without Nash's 
resummation and $3.17 < N_c <  3.24$ with Nash's resummation) are close to $N_c=7.2/2=3.6$ obtained in  \cite{Khveshchenko08}
and $N_c=3.52$ \cite{LiuLC09}.
%
These results for $\al_c$ would not be compatible with the semimetallic behaviour of graphene observed experimentally \cite{Eliasetal11} if we were to compare
them with the bare coupling constant $\al \approx 2.2$ in clean suspended graphene. We may however argue that the
renormalization of the Fermi velocity observed in \cite{Eliasetal11} would rather be compatible with a coupling constant of about $0.73$ which is indeed smaller than all 
of the above values theoretically obtained for $\al_c$. 

Nevertheless, the results \cite{Khveshchenko01,GorbarGMS02,Khveshchenko08,GamayunGG09} were then
criticized for not properly taking into account dynamical screening of interactions and/or wave function and/or velocity renormalizations.
Recent attempts to better take into account (some or all of) these effects at the level of SD equations led to larger values: $3.2 < \al_c < 3.3$ \cite{WangL12}, 
$\al_c = 7.65$ \cite{Popovici13} and $\al_c = 3.1$ \cite{Gonzalez15}. As discussed in \cite{Katanin15}, where the value $\al_c = 3.7$ was obtained using different methods,
the result $\alpha_c=3.1$ seems to be the most reliable within the Schwinger-Dyson approach.  
Such large values are well above the bare value of $\al$ and therefore compatible with the semi-metallic ground state observed experimentally.

\section{Conclusion} 
\label{sec:conclude}

The critical behaviour of RQED$_{4,3}$ was studied up to NLO on the basis of a correspondence with QED$_3$
 providing high precision estimates of the critical coupling constant, $\al_c$, 
and critical fermion flavour number, $N_c$, for graphene at the fixed point.
Dynamical screening and wave function renormalization were fully taken into account (there is no velocity renormalization at the fixed point).
We found that $\al_c \sim \Ord(1)$, so that $\al_c \gg \al_{\text{QED}}$ and, at the fixed point, graphene is deep in the semimetallic phase in qualitative 
agreement with experiments in actual samples. The striking feature of our results is that our values of $\al_c$ and $N_c$
are in good {\it quantitative} agreement with results obtained in the non-relativistic limit \cite{LealK03,Khveshchenko08,Drut08,LiuLC09,GamayunGG09,BuividovichP13}
 including lattice simulations \cite{Drut08,BuividovichP13}. 
Such an agreement between the two extreme limits, $v/c \ra 0$ and $v/c \ra 1$, 
seems to suggest that the study of the fixed point is not only of academic interest
and that our model may be an efficient effective field theory model in describing some of the features of actual planar condensed matter physics systems \footnote{Let us recall that, 
in a completely different context, a rather good agreement was found for the value of the interaction correction coefficient to the
optical conductivity of graphene in the two extreme limits $v/c \ra 0$ and $v/c \ra 1$, see Refs.~\cite{Teber12+KotikovT13}, \cite{TeberK14} as well as \cite{TeberK16} for a short review.}.  
An interesting and difficult challenge would then be to extend all of these computations to arbitrary values of $v/c$. We leave this task for future work.

\acknowledgments
One of us (A.V.K.) was supported by RFBR grant 16-02-00790-a.
Financial support from Universit\'e Pierre et Marie Curie and CNRS is acknowledged.


\end{document}